\documentclass[floatfix,superscriptaddress,reprint,aps,pra,longbibliography]{revtex4-1} 
\usepackage{amsfonts,amsmath,amssymb}
\usepackage{color}
\usepackage{commath}%
\usepackage{esint}%
\usepackage[T1]{fontenc}
\usepackage{graphicx}
\usepackage[utf8]{inputenc}
\usepackage{microtype}
\usepackage{times}
\usepackage[textsize=footnotesize,obeyFinal]{todonotes}
\usepackage{xspace}
\usepackage[colorlinks=false]{hyperref}
\usepackage{hypernat}
\usepackage[todos]{CMImacros}

\graphicspath{{./figures/}} 

\def\mP{\mathcal{P}}
\def\mQ{\mathcal{Q}}

\def\wint{\ensuremath{\text{W/cm}^2\,}}

\def\pD#1#2{\frac{\partial}{\partial #2}#1}
\def\pDD#1#2{\frac{\partial^2}{\partial {#2}^2}#1}

\def\FF{\left(\frac{1}{3}f_1+\frac{1}{2}f_2\right)}
\def\HH{\left(\frac{1}{3}f_1-\frac{1}{2}f_2\right)}

\newlength{\figwidth}
\newcommand{\cfeldesy}{\affiliation{Center for Free-Electron Laser Science, Deutsches
      Elektronen-Synchrotron DESY, Notkestrasse 85, 22607 Hamburg, Germany}}%
\newcommand{\uhhcui}{\affiliation{The Hamburg Center for Ultrafast Imaging, Universität Hamburg,
      Luruper Chaussee 149, 22761 Hamburg, Germany}}%
\newcommand{\uhhphys}{\affiliation{Department of Physics, Universit\"at Hamburg, Luruper Chaussee
      149, 22761 Hamburg, Germany}}%
\newcommand{\mpi}{\affiliation{Max Planck Institute for the Structure and Dynamics of Matter, 22761
      Hamburg, Germany}}%
\newcommand{\whu}{\affiliation{Department of Physics, Wuhan University, 430072 Wuhan, China}}%
\newcommand{\desy}{\affiliation{Deutsches Elektronen-Synchrotron DESY, Notkestrasse 85, 22607
      Hamburg, Germany}}%
\newcommand{\peking}{\altaffiliation[Present~Address:~]{School of Physics, Peking University,
      Beijing, China.}}%

\begin{document}
\title{Acoustic funnel and buncher for nanoparticle injection}%
\author{Zheng Li}\email{zheng.li@desy.de}\peking\mpi\cfeldesy%
\author{Liangliang Shi}\thanks{Present address: Paul Scherrer Institut PSI, CH-5232 Villigen, Switzerland}\desy%
\author{Lushuai Cao}\thanks{Z. Li, L. Shi and L. Cao contributed equally to this work.}\uhhcui%
\author{Zhengyou Liu}\whu%
\author{Jochen Küpper}\email{jochen.kuepper@cfel.de}\homepage{https://www.controlled-molecule-imaging.org}\cfeldesy\uhhcui\uhhphys%
\begin{abstract}\noindent%
   Acoustics-based techniques are investigated to focus and bunch nanoparticle beams. This allows to
   overcome the prominent problem of the longitudinal and transverse mismatch of particle-stream and
   x-ray-beam size in single-particle/single-molecule imaging at x-ray free-electron lasers (XFEL).
   It will also enable synchronized injection of particle streams at kHz repetition rates.
   Transverse focusing concentrates the particle flux to the size of the (sub)micrometer x-ray
   focus. In the longitudinal direction, focused acoustic waves can be used to bunch the particle to
   the same repetition rate as the x-ray pulses. The acoustic manipulation is based on simple
   mechanical recoil effects and could be advantageous over light-pressure-based methods, which rely
   on absorption. The acoustic equipment is easy to implement and can be conveniently inserted into
   current XFEL endstations. With the proposed method, data collection times could be reduced by a
   factor of $10^4$. This work does not just provide an efficient method for acoustic manipulation
   of streams of arbitrary gas-phase particles, but also opens up wide avenues for acoustics-based
   particle optics.
\end{abstract}
\maketitle

\noindent%
X-ray free-electron lasers enable single-particle and single-molecule imaging by x-ray
diffraction~\cite{Spence:PTRSB369:20130309}, due to the unprecedented brightness and femtosecond
pulse duration. As the particle stream enters the vacuum chamber, transverse expansion is inevitable
for freely moving particles due to the pressure difference. At present, one of the key bottlenecks
in single-particle imaging at XFELs is the large size of aerodynamically focused particle streams,
often of a few tens of micrometers~\cite{Hantke:NatPhoton8:943, Awel:JAC51:133} compared to the
small size of the 100~nm-diameter x-ray beam. Furthermore, in the longitudinal direction the
particles passing between the pulses are also not intercepted. This mismatch results in low sample
delivery efficiency, only about one in $10^{12}$ particles are intercepted in the case of a 100~\um
particle beam moving at 100~m/s across a 100~nm x-ray beam at a 1~kHz repetition rate. As a result,
many samples, which are often precious, are wasted, and days of data collection are often required
in order to obtain only a few hundred or perhaps thousand high-quality diffraction patterns at an
x-ray pulse repetition rate of some kHz, whereas $>\!100000$ patterns are required for
atomic-resolution imaging~\cite{Barty:ARPC64:415}.

Different means to enhance the interception rate of particles by the x-ray pulses through transverse
focusing are considered, such as improved aerodynamic collimation~\cite{Kirian:SD2:041717,
   Roth:407877, Horke:JAP:123106} or the focusing with laser traps~\cite{eck15:064001}. Furthermore,
bunching~\cite{Meerakker:CR112:4828}, \ie, longitudinal focusing, with spatial periods that match
the repetition rate of x-ray pulses could be utilized to further improve sample use. Suppose the
particles stream was transversely compressed to 1~\um and bunched to millimeter size with the same
frequency as the repetition rate of x-ray pulses in the longitudinal direction: compared to the
typical parameters given above, data collection time and sample use would be reduced by a factor of
$10^6$.

\begin{figure}[t]
   \includegraphics[width=0.75\linewidth]{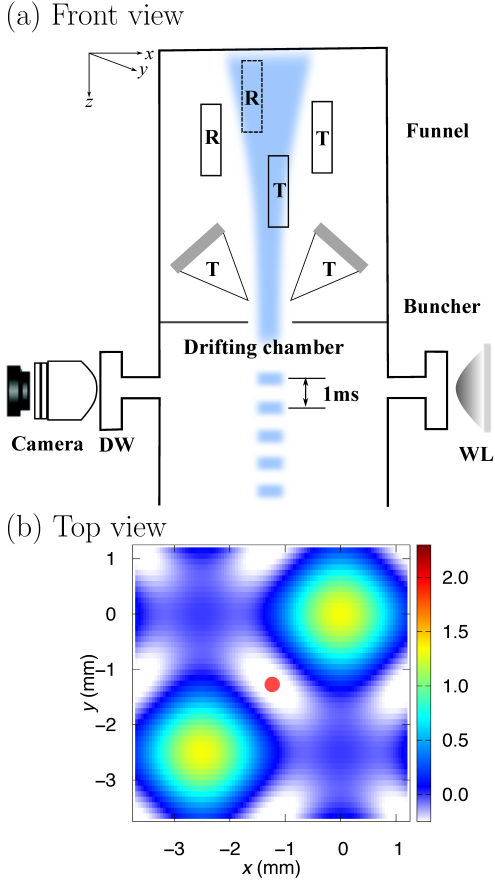}
   \caption{Front view (a), with three-dimensional perspective, of a sketch of the setup, including
      the configuration of the acoustic funnel and buncher as well as the detection window (DW),
      which is equipped with white light (WL) illumination and camera. The acoustic funnel is formed
      by two orthogonal standing wave resonators in the $x,y$ directions, each consisting of a
      transducer (T) and a reflector (R), and the particle flow is along $z$ direction. A top view
      of the created acoustic potential is shown in (b). The coordinates refer to the center of the
      mechanical setup of transducers and reflectors. Potential minima are characterized by an oval
      shape and the center of one is marked by a red dot at $-(\lambda/4,\lambda/4)$, corresponding
      to the position where the focusing experiment is performed. Below the funnel, the acoustic
      buncher is formed by two tilted transducers that emit synchronized acoustic waves. The
      acoustic wave is transversely focused by a conical cavity with a pinhole~\cite{kaz06:4560}.
      The upper chamber is filled with helium gas, at a pressure of $10^{-3}$~mbar, as the acoustic
      coupling medium. The particle stream enters from the top and moves downward.}
   \label{fig:setup}%
\end{figure}

Here, we propose that the longitudinal and transverse manipulation of the particle stream can be
realized by an acoustic funnel and a buncher as sketched in \autoref{fig:setup}. In gas flows, the
deviation from continuum behavior is quantified by the Knudsen number, $\Knud=\Lambda/H$, where
$\Lambda$ is the mean free path and $H$ is a characteristic length scale, which can be taken as the
width between transducer and reflector. $\Knud>10$ corresponds to ballistic molecular behavior of
free molecular flow, $0.1\leq{}\Knud\leq10$ is known as the transition regime, and for
$\Knud\lesssim0.1$ a continuum hydrodynamic description is possible. We focus on the case of
$\Knud<0.1$, for which the conventional picture of acoustic waves in continuum media is
valid~\cite{had02:802}. For helium gas at $T=5$~K the mean free path is
$\Lambda=k_BT/\sqrt{2}\pi\sigma^2p=2$~mm, with the size of the helium atom $\sigma=280$~pm and the
pressure $p=10^{-3}$~mbar, or similarly for $p=5\times10^{-2}$~mbar at room temperature. The width
of the standing wave resonator is $H=(n+1/2)\lambda=2.75$~cm with an acoustic wave of wavelength
$\lambda=5$~mm and frequency $\nu=26$~kHz. In the following, we present the theory for the
transverse and longitudinal manipulation with standing and traveling acoustic waves, respectively.

As illustrated in \autoref{fig:setup}, the acoustic funnel is made of two orthogonal half-wavelength
cavities formed by transducers and specular reflectors in the transverse direction. The two 1D
cavities are set up to overlap in the center of the particle beam. Since the focusing in the
transverse $x,y$ directions is similar, we firstly consider the Gor'kov potential $U(x,y;t)$ for focusing in the
$x,y$ direction~\cite{gor62:88,Oberti:JASA121:778,li19:013702}
\begin{eqnarray}
  \label{eq:FGor}
  U &=& \frac{16\pi
    R^3I}{c}\left[\frac{1}{3}f_1(\cos^2{kx}+\cos^2{ky}+2\cos{kx}\cos{ky})\right.\nonumber\\
    &\times&\left.\sin^2{\omega t}
    -\frac{1}{2}f_2(\sin^2{kx}+\sin^2{ky})\cos^2{\omega t}\right]
\end{eqnarray}
with $f_1=1-(\rho{}c^2)/(\rho_0{}c_0^2)$ and $f_2=2(\rho_0-\rho)/(2\rho_0+\rho)$. $I$ and $k$ are
the intensity and wave number of the acoustic field, respectively, $R$ is the radius of the
particle, $c$ and $\rho$ are the speed of sound in and the density of the coupling medium, and $c_0$
and $\rho_0$ are the speed of sound in and the density of the particle. Due to the fast velocity of
the nanoparticles, we keep the form of Gor'kov force with temporal modulation~\cite{gor62:88}. As
will be shown below, the exact form of static Gor'kov force relies on the condition that the
characteristic frequency of particle motion has to be much lower than that of the acoustic wave,
such that the particles have stable trajectories, and this condition can be well fulfilled in our
scheme.

We assume mass and radius of the particle as $m=3\times{10^{-21}}~\text{kg}$ and $R=100$~nm, which
resembles typical biological sample particles, such as virus particle. \eqref{eq:FGor} corresponds
to the force from the potential of an eigen mode that has a minimum at the center of the cavity
$r=0$~\cite{and15:014317, bar85:928, rae11:014317}. Assuming the particle has, at least, one
symmetry axis and the longitudinal motion is parallel to that axis, there is no deflecting force in
the transverse direction~\cite{lan59}. Thus the Brownian motion is the dominant mechanism of
transverse dispersion of the particle beam. Denoting the transverse velocity as $v_y=\dot{y}$, the
equation of motion for the Brownian motion in Gor'kov potential is
\begin{equation}
   \begin{aligned}
      \label{eq:eom}
      & m\dot{v}_y+\beta v_y=F_B(t)+F_G(y,t) \\
      & v_y(0)=0\,, y(0)=0 \,,
   \end{aligned}
\end{equation}
where $F_B(t)$ is the force of Brownian collision and $F_G(y,t)$ is the Gor'kov force.
For low pressure, $p\lesssim10^{-3}$~mbar, helium as the coupling medium, and a Knudsen number close
to the transition regime, the friction coefficient $\beta$ can be expressed as
\begin{equation}
  \label{eq:beta}
  \beta = 4\pi R^2\rho\sqrt{\frac{2k_BT}{m_a}}
  \,.
\end{equation}
We can linearize the Gor'kov force around $-(\lambda/4,\lambda/4)$ as
\begin{eqnarray}
     F_G(y,t) &=& -\frac{16\pi{I}R(kR)^2}{c}\left[\FF-\cos{2\omega t}\right.\nonumber\\
       &\times& \left.\HH\right]y \nonumber\\
      &=& -Gy\left(1+\frac{H}{G}\cos{2\omega{t}}\right)
      \,.
\end{eqnarray}
The motion in the $x$-direction is the same, since the linearized Gor'kov force $F_G(x,t)$ can be
obtained by replacing $y$ with $x$.
The oscillating term in the Gor'kov force that is proportional to $\cos{2\omega{t}}$ can possibly
induce parametric resonances and drive particles away from the equilibrium position of the
potential. However, it can be shown that the parametric resonance can be safely avoided in our case,
due to a large difference between the frequencies of particle oscillation and acoustic wave:
Rewriting \eqref{eq:eom} approximately in the form of a Mathieu equation
\begin{equation}
   \label{eq:par}
   \ddot{y}+\frac{\beta}{m}\dot{y}+\frac{G}{m}y(1+\frac{H}{G}\cos{2\omega{t}})=0
   \,.
\end{equation}
and denoting $\Omega=\sqrt{{G}/{m}}$ as the characteristic frequency of particle oscillation, the
particle trajectory is found as
\begin{equation}
   \begin{aligned}
      y(t) =& e^{-\frac{\beta{t}}{2m}}
      \left\{C_1\mathcal{C}\left[\left(\frac{\Omega}{\omega}\right)^2-\left(\frac{\beta}{2m\omega}\right)^2,
            -\frac{H}{2G}\left(\frac{\Omega}{\omega}\right)^2,
            \omega{t}\right]\right.\\
      &+ \left.C_2\mathcal{S}\left[\left(\frac{\Omega}{\omega}\right)^2-\left(\frac{\beta}{2m\omega}\right)^2,
            -\frac{H}{2G}\left(\frac{\Omega}{\omega}\right)^2,
            \omega{t}\right]
      \right\}
      \,,
   \end{aligned}
\end{equation}
where $\mathcal{C}(a,q,\nu)$ and $\mathcal{S}(a,q,\nu)$ are even and odd Mathieu functions.
\begin{figure}
   \includegraphics[width=\linewidth]{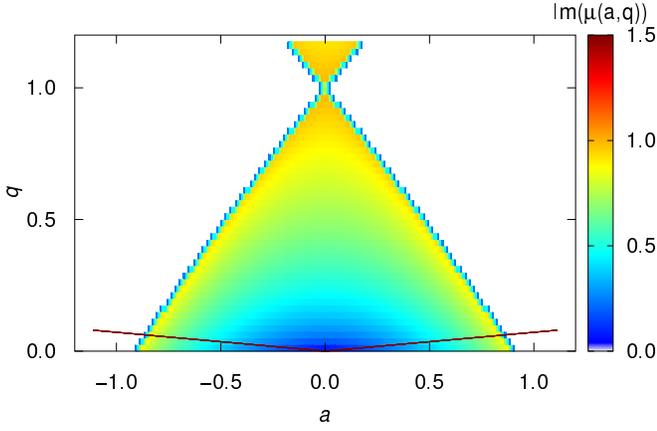}
   \caption{Stability diagram according to \eqref{eq:par} with $\beta=0$. The region with a purely
      imaginary characteristic exponent of ${\rm Im}[\mu(a,q)]>0$ permits stable trajectories
      according to the Mathieu equation, see color map, and the unstable region is left white. The
      parameters for our case correspond to the solid lines $a=(\Omega/\omega)^2=(2G/H)q$.}
   \label{fig:stable}%
\end{figure}
Rigorous theory of Mathieu equations gives the stable regime of the particle trajectory with
$\beta=0$~\cite{tim06:mathieu, pau90:531}, see \autoref{fig:stable}. In this parameter space the
particles oscillate transversely with limited amplitudes that do not grow exponentially. A wide
range of ratios between particle-oscillation and acoustic-wave frequencies provide stable
trajectories. The required condition can be conveniently fulfilled even without friction, \eg, in
our case $(a,q)\simeq(0.11,4.4\times{10^{-4}})$. Variational analysis demonstrated that the friction
can further widen the permitted stable regime according to the Mathieu
equation~\cite{hsi77:1147,hsi80:722}, since it physically suppresses the oscillation amplitude of
particle trajectory.

Computations following \eqref{eq:par} show converging trajectories to the center of the harmonic
potential. In the absence of parametric resonances, the particle trajectories must converge to the
focused area. Similar to the case of a pure harmonic potential the temporal factor in the Gor'kov
force could be approximately integrated out~\cite{gor62:88}. Based on the stability analysis, the
particle's velocity is
\begin{equation}
  v_y(t)=-\frac{\beta}{m}y+\frac{1}{m}\int_0^t F_B(\zeta)d\zeta + \frac{1}{m}F_G(y)t
  \,.
\end{equation}
The corresponding Fokker-Planck equation~\cite{fok1914, pla1917, orn19:96} for the transverse
distribution of the particles $f(y,y_0,t)$ can thus be obtained, for the linearized Gor'kov force,
as
\begin{equation}
  \label{eq:fp}
  \pD{f}{t}=\frac{G}{\beta}\pD{(yf)}{y}+D\pDD{f}{y}
  \,.
\end{equation}
From the Fokker-Planck equation, the temporal evolution of the transverse particle positions are
obtained as
\begin{equation}
   \label{eq:fsol}
   \begin{aligned}
      f(y,y_0,t) =& \left[ \frac{G}{2\pi\beta D(1-e^{-\frac{2G}{\beta}t})} \right]^{1/2} \\
      &\times \exp\left[-\frac{G}{2{\beta}D}\frac{\left(y-y_0e^{-\frac{G}{\beta}t}\right)^2}{1-e^{-\frac{2G}{\beta}t}}\right]
      \,.
   \end{aligned}
\end{equation}
This yields the minimal width of the particle stream as
\begin{equation}
   \label{eq:minimal_width}
   w_{\text{min}}=\sqrt{\frac{2\ln{2}k_BT}{G}}
   \,.
\end{equation}
Given an initial width $w_0$, the transverse distribution function $f(y,y_0,t)$ in \eqref{eq:fsol}
can be convoluted as
\begin{equation}
   \begin{aligned}
      f(y,t) &= \int_{-\infty}^{\infty}f(y,y_0,t)w(y_0)dy_0 \\
      w(y_0) &= \sqrt{\frac{\ln2}{{\pi}w_0^2}} e^{-\ln2 \, y_0^2 / w_0^2} \,,
   \end{aligned}
\end{equation}
which gives the temporal evolution of particle stream
\begin{align}
  f(y,t) =& \mP\sqrt{\frac{\pi}{{\mQ}e^{-\frac{2G}{\beta}t}+\sqrt{\ln{2}/w_0^2} }} \\
          &\times\exp\left[-\left(\mQ-\frac{\mQ^2e^{-\frac{2G}{\beta}t}}{{\mQ}e^{-\frac{2G}{\beta}t}+\sqrt{\ln{2}/w^2_0}}\right)y^2\right]
            \nonumber \, ,
\end{align}
where $\mP(t)=\sqrt{G\ln{2}/(2\pi{}w_0^2\beta{D}(1-e^{-{2Gt}/{\beta}}))},$ and
$\mQ(t)=G/(2\beta{D})\cdot1/(1-e^{-{2Gt}/\beta}).$

\begin{figure}
   \includegraphics[width=\linewidth]{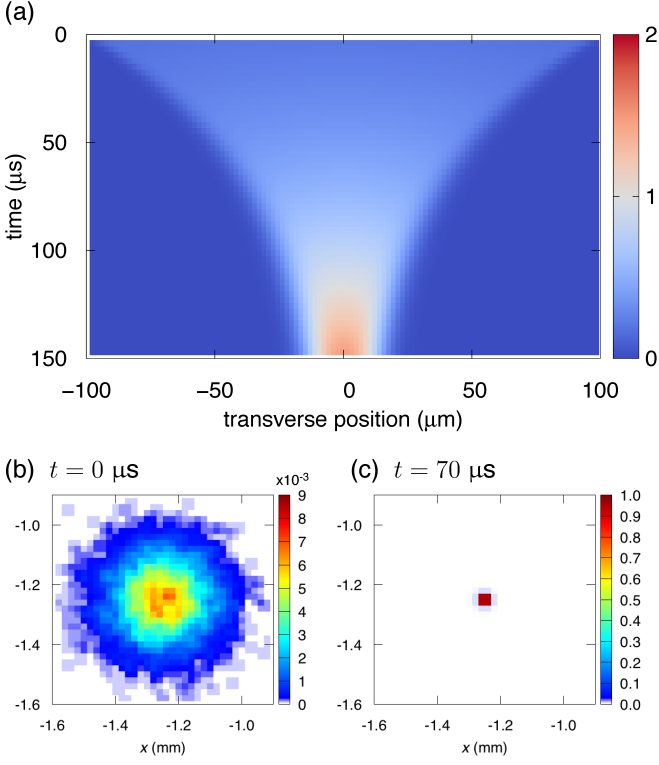}
   \caption{(a) Temporal evolution of a particle distribution with an initial width (waist)
      $w_0=100~\um$ in an acoustic wave with frequency $\nu=26$~kHz and an intensity of $I=1~\wint$.
      The final width is consistent with the minimal width $w_{\text{min}}=7.5~\um$, determined from
      \eqref{eq:minimal_width}. Particle number density distributions are plotted at (b) 0~\us and
      (c) 70~\us from numerically simulated dynamics using the 2D potential with explicit time
      dependence in \eqref{eq:FGor}. The root mean squared deviation from the center
      $(-\lambda/4,-\lambda/4)$ is 100~\um and 6~\um at 0~\us and 70~\us, respectively.}
   \label{fig:result_init_w}
\end{figure}

The temporal evolution obtained for $w_0=100~\um$ is presented in \autoref{fig:result_init_w}.
The particle beam is transversely compressed to a width of $6~\um$, approaching the size of the XFEL
beam.
We show the temporal evolution of particle number density distribution determined from \eqref{eq:fsol} in \autoref{fig:result_init_w}(a), and from numerical simulations in \autoref{fig:result_init_w}(b) and (c).

The acoustic buncher relies on the period force imposed by the traveling wave resulting from tilted
transducers, see \autoref{fig:setup}. Suppose the two transducers radiate synchronously with the
same phase, then the transverse force is zero and only a force in the longitudinal direction
remains. In our case, the particles move with a longitudinal velocity of $v_z\sim100$~m/s, and the
buncher imposes a force field that has sufficiently short longitudinal interaction length, \ie, the
particle transit time $\Delta{t}=l/v_z$ is much shorter than the period of the acoustic wave. Since
the acoustic pressure variation does not affect the particle for a full cycle, the particle only
experience a transient force. This leads to an acoustic force that is proportional to the first
order of the sinusoidal modulation of the plane acoustic wave. Assuming the acoustic pressure to be
$p=p_0\sin(\vec{k}\cdot\vec{r}-\omega t+\phi_0)$, it takes a form
\begin{equation}
  p=p_0\sin\left[\vec{k}\cdot\left(\vec{r}_0(t)+\vec{R}\right)-\omega{t}+\phi_0\right]
  \,,
\end{equation}
for $\vec{r}=\vec{r}_0(t)+\vec{R}$ on the surface of a particle at position $\vec{r}_0(t)$ with
radius R, where $\vec{k}$ is the wave vector and $\phi_0$ is an arbitrary phase. Under this
assumption, the acoustic force exerted on the particle is
\begin{equation}
   \begin{aligned}
      f_z =& \oiint p dS = \int_0^{2\pi}d\phi \int_0^{\pi}d\theta\sin\theta R^2p_0 \\
      &\times \sin\left[ kR\cos\theta-\omega{t} +\vec{k}\cdot\vec{r}_0(t)+\phi_0 \right] \\
      =& \frac{4\pi R\sin{kR}}{k}\sin\left(\omega{t}-\vec{k}\cdot\vec{r}_0(t)-\phi_0\right)
      \,.
   \end{aligned}
\end{equation}
Because of the narrow width of the interaction area of the buncher, and the $\lesssim100$~nm size of
the particles, \ie, $kR\ll1$, the particle scattering effect is suppressed, and the force can be
further approximated as $f_z\simeq p_0S\sin(\omega{t}-\theta_0)$, where $\theta_0$ is a constant
phase factor, left to be chosen, and $S$ is the surface area of the particle. In general, we write
the effective longitudinal force as $f_z=F\sin\omega{t}$.

Considering an acoustic wave with $\nu=1$~kHz and an induced relative pressure variation of
$\Delta{}p=2\times10^{-4}$~mbar, well below the pressure in the chamber, we obtain a force
$F=10^{-4}$~pN. The particles experience a periodic velocity modulation with respect to their
entrance time into the interaction region of length $d$, yielding
$\frac{1}{2}mv^2-E_1\approx{}Fd\sin\omega{t_1}$ with the particle velocity $v_1$ and kinetic energy
$E_1=\frac{1}{2}mv_1^2$ at the entrance of the interaction region. The length $d$ of the
particle-wave-interaction region is chosen such that the particle experiences the force over only
$1/10$ of the acoustic wave period. A conical cavity with a pin-hole can focus the acoustic wave to
a length on the order of $\lambda/40$ in the near field~\cite{kaz06:4560}. Considering the
wavelength of the 1~kHz acoustic wave, $\lambda=11$~cm, we choose the length of interaction region
to be $d=1$~cm, which is experimentally feasible. Thus we have approximately
$v=v_1\left[1+(Fd/2E_1)\sin\omega{t_1}\right].$

Assuming particles drift for a distance $l$ after leaving the interaction region and arrive at the
end of the buncher at time $t_2$, we have
\begin{equation}
   t_2=t_1+\frac{l}{v} \simeq t_1+\frac{l}{v_1}\left(1-\frac{Fd}{2E_1}\sin{\omega}t_1\right)
   \,.
\end{equation}
With an initial number density $n_1$ and the continuity condition $n_2dt_2=n_1dt_1$, the modulated
number density $n_2$ at $t_2$ can be expressed as
\begin{align}
  n_2 &= n_1+\sum_{k=1}^{\infty}a_k\cos\left[k(\omega{t_2}-\Theta)\right]+b_k\sin\left[k(\omega{t_2}-\Theta)\right]\nonumber\\
  \text{with} \\
  a_k &= \frac{n_1}{\pi}\int_{\Theta-\pi}^{\Theta+\pi}\cos\left[k(\omega{t_1}-X\sin{{\omega}t_1})\right]d(\omega{t_1})\nonumber\\
      &= 2n_1J_k(kX)\nonumber\\
  b_k &= 0 \qquad \text{for}\,\,k=1,2,\ldots \,, \nonumber
\end{align}
where $\Theta=l\omega/v_1$, $X=Fdl\omega/(2E_1v_1)$, and $J_k(x)$ is the Bessel function of $k$-th
order. We consider the fundamental harmonic
\begin{equation}
   n_2=n_1+2n_1J_1(X)\sin({\omega}t_1-\Theta)
   \,.
\end{equation}
The degree of bunching is determined by the bunching parameter $X$. The frequency of the traveling
wave can be conveniently set as the repetition rate of x-ray pulses.

After the particle stream passes the interaction region of length $d$ it can continue into the next
chamber, see \autoref{fig:setup}, and drifts for a distance $l$ to the interaction point. Assuming
$I=1~\wint$, $\nu=1$~kHz, $d=1$~cm, $v_1=100$~m/s, and that the cavities are tilted by $\Psi=\pi/3$,
the degree of bunching is maximized as the Bessel function $J_1(X)$ reaches its maximum at
$X\simeq{1.8}$, which corresponds to a drift length $l=87$~cm.

We numerically simulate the bunching process using particle tracing methods~\cite{elegant2000}.
\begin{figure}
   \centering%
   \includegraphics[width=0.8\linewidth]{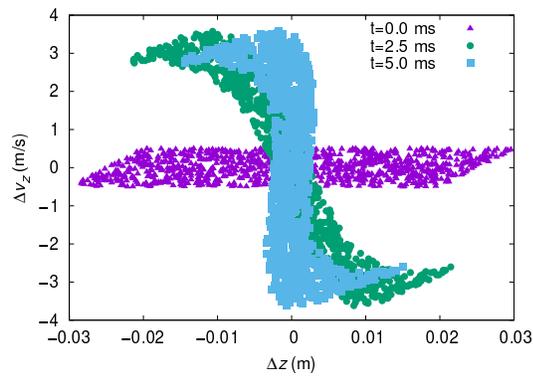}
   \caption{The calculated longitudinal phase-space distribution of the particles is given at the
      entrance of the buncher, $t=0.0$~ms, and that in the detection region at $t=5.0$~ms as well as
      an intermediate time $t=2.5$~ms demonstrating the phase-space rotation; all distributions
      relative to the phase-space position of the synchronous particle.}
   \label{fig:buncher}
\end{figure}
In the simulation, the buncher is operated such that a 6~cm long packet of molecules with a
longitudinal velocity of 100~m/s and a velocity spread of 1~m/s enters the acoustic buncher. The
impulse by a force of $10^{-4}$~pN acting on particle of $3\times 10^{-21}$~kg for
$\Delta{t}=d/v_1{\simeq}0.1$~ms can modulate the velocity by $\Delta{v}\simeq 3.3$~m/s. This can be
used as the criterion to choose the acoustic pressure, since the modulation must be similar to that
of the velocity spread of the particle beam. The calculated distribution at $t=5$~ms, the time at
which the longitudinal spatial focus is obtained downstream of the buncher, is shown in
\autoref{fig:buncher}. The longitudinal phase space distribution is relative to the position in
phase space of the ``synchronous particle''~\cite{Meerakker:CR112:4828}. In the particular situation
depicted in \autoref{fig:buncher}, the molecular packet has a longitudinal focus with a length of
about 3~mm some 53~cm after the end of the buncher. The longitudinal focal length is consistent with
our simplified model with a single velocity and infinitely short interaction region.

We have proposed an acoustic method to manipulate and compress particle streams by transverse and
longitudinal focusing, which enables high-efficiency particle delivery, for instance, for
single-particle diffractive imaging experiments with sub-\um-focus x-ray beams. This can
substantially reduce the data collection time in such XFEL based imaging experiments. The effective
manipulation of particle streams based on acoustic waves could be applied to wider scope of
molecular beam experiments, such as matter-wave-interference with large
molecules~\cite{Arndt:NatPhys10:271} as well as applications to fast highly collimated
beams~\cite{Kirian:SD2:041717}. Furthermore, this work does not just provide an efficient method for
acoustic manipulation of gas-phase-particle streams, but also sheds light on the application of the
vast particle-optics techniques from accelerator physics to the field of acoustics, \eg, such as
particle bunching by the traveling wave from analogues to iris-loaded waveguides.

The authors gratefuly acknowledge stimulating discussions with R.\ J.\ Dwayne Miller, Oriol
Vendrell, Nikita Medvedev, Sheng Xu, Ludger Inhester, and Henry N.\ Chapman.

This work has been supported by a Peter Paul Ewald Fellowship of the Volkswagen Foundation, by the
European Research Council under the European Union's Seventh Framework Programme (FP7/2007-2013)
through the Consolidator Grant COMOTION (ERC-614507-Küpper), by the Clusters of Excellence ``Center
for Ultrafast Imaging'' (CUI, EXC~1074, ID~194651731) and ``Advanced Imaging of Matter'' (AIM,
EXC~2056, ID~390715994) of the Deutsche Forschungsgemeinschaf, and by the Helmholtz Gemeinschaft
through the ``Impuls- und Vernetzungsfond''.

\bibliography{bremse}
\onecolumngrid
\end{document}